\documentclass[a4paper]{article}

\usepackage{INTERSPEECH2020}
\usepackage{url}
\usepackage{cite}

\title{Listen to What You Want: Neural Network-based Universal Sound Selector}
\name{Tsubasa Ochiai, Marc Delcroix, Yuma Koizumi, Hiroaki Ito, Keisuke Kinoshita, Shoko Araki}
\address{
  NTT Corporation, Japan}
\email{tsubasa.ochiai.ah@hco.ntt.co.jp}

\begin{document}

\maketitle
\begin{abstract}
Being able to control the acoustic events (AEs) to which we want to listen would allow the development of more controllable hearable devices. This paper addresses the AE sound selection (or removal) problems, that we define as the extraction (or suppression) of all the sounds that belong to one or multiple desired AE classes.
Although this problem could be addressed with a combination of source separation followed by AE classification, this is a sub-optimal way of solving the problem.
Moreover, source separation usually requires knowing the maximum number of sources, which may not be practical when dealing with AEs.
In this paper, we propose instead a universal sound selection neural network that enables to directly select AE sounds from a mixture given user-specified target AE classes.
The proposed framework can be explicitly optimized to simultaneously select sounds from multiple desired AE classes, independently of the number of sources in the mixture.
We experimentally show that the proposed method achieves promising AE sound selection performance and could be generalized to mixtures with a number of sources that are unseen during training.

\end{abstract}

\section{Introduction}

In our daily lives, we are constantly immersed in acoustic scenes or environments that are composed of mixtures of acoustic events (AEs), which are often called polyphonic sounds~\cite{cakir2015polyphonic,mesaros2016metrics}.
Depending on the situation, AEs can provide critical information about our surroundings, e.g., klaxons when crossing a street. They can also be annoying disruptions of our concentration, e.g., similar klaxons when working at home with an open window.
Our daily lives could be greatly improved if we could develop hearable devices that select the AEs that we want to listen to depending on the situation. 
Toward this goal, we define two problems: 1) AE sound selection and 2) AE sound removal.
The AE sound selection problem consists of the extraction of the one or multiple desired AE sounds from the mixture, given the user-specified target AE classes.
When multiple desired AE classes are selected, we output a signal that consists of the sum of all the AEs from these classes.
In this paper, we focus on the AE sound selection problem, and briefly discuss the related AE sound removal problem, which consists of the suppression of the undesired AE sounds from the mixture.
To the best of our knowledge, this paper is the first attempt 
that explicitly tackles these two problems. 

Thus far, research on AE processing has focused on AE detection/classification (AED) and AE separation (AES): separating a mixture of AEs into individual AEs.
AED detects the boundaries and identifies the classes of each AE in a recording, which may have many practical applications, e.g., for surveillance~\cite{clavel2005events,atrey2006audio}.
AEDs have been extensively researched, and rapid progress in the field has been fueled by the recent DCASE challenge series~\cite{mesaros2017dcase} and the release of large datasets~\cite{fonseca2018general,gemmeke2017audio}.
Since AED enables only to detect sounds without extracting them from the acoustic scene, they are thus insufficient to solve the AE selection problem that we target. 

Recently, the separation of AEs using neural networks to output signals for each AE in a mixture has received increased interest~\cite{kavalerov2019universal,tzinis2020improving}.
For example, the utterance-level permutation invariant training (uPIT)~\cite{kolbaek2017multitalker} framework, which was initially developed for speech processing, has been extended for the separation of AEs~\cite{kavalerov2019universal}.
Although such an approach could separate the AEs, it has two main drawbacks. 
First, the number of AEs that it can separate is fixed by the number of network outputs.
In the case of speech separation, it would be reasonable to fix the maximum number of outputs to two or three since in practice it is rare that more relevant speakers speak within an audio segment.
However, the number of AEs can be substantially larger in an acoustic scene. Training a PIT-based separation network for a larger number of outputs is challenging because of the exponential increase in the number of permutation possibilities in the PIT loss.
Another issue is the global permutation ambiguity at the separation's output, i.e., it is ambiguous which output corresponds to which source AE.
Therefore, AES alone also fails to solve the AE selection problem.

One way to achieve AE sound selection would be to combine AES and AED.
For example, with a cascade combination of AES and AED obtained by performing AED after a PIT based AES~\cite{dcase2020}, the desired AEs could be picked up from the outputs.
This approach has three limitations.
First, it inherits the drawbacks of PIT-based separation in terms of the maximum number of AEs it can manage.
Second, the AED performance after separation may not be optimal because of the possible processing artifacts or remaining undesired sounds in the separation output.
Finally, we need to perform AED on all the separation outputs, which may become computationally expensive when dealing with many possible simultaneous AEs.

In this paper, we propose a neural network-based AE sound selection approach, called Sound Selector, which directly extracts the desired AE sound from a mixture of AEs given a one-hot vector representing the class of interest.
Since the Sound Selector outputs a single signal independently of the number of AE sounds in the mixture, it does not require knowing the maximum number of AEs like a PIT-based AES.
Moreover, the Sound Selector is explicitly optimized based on an objective function for AE sound selection. 
Finally, we extend the Sound Selector to simultaneously extract AEs from multiple AE classes, making it computationally efficient, which would be important for future extensions of the approach to real-time processing on hearable devices.


\section{Proposed method}

In this paper, we define two novel concepts, i.e., AE sound selection and AE sound removal.
In the following for discussion simplicity, we focus on AE sound selection and describe our proposed framework that extracts (selects) the desired AE sounds from the observed mixture (Figure~\ref{fig:example} for an example), given the user-specified target AE classes.

\begin{figure}[t]
  \centering
  \includegraphics[width=0.75\linewidth]{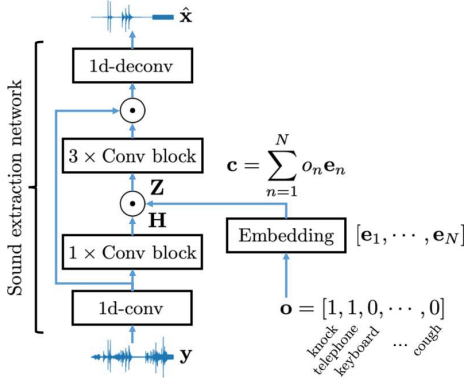}
  \vspace{-2mm}
  \caption{Overview of proposed Sound Selector architecture.}
  \label{fig:network}
  \vspace{-3mm}
\end{figure}

\subsection{Sound selection framework}

Let $\mathbf{y} \in \mathbb{R}^{T}$ be the $T$-length time-domain waveform of the observed mixture.
We assume that target-class vector $\mathbf{o} = [ o_{1}, \cdots, o_{N} ]^{\mathsf{T}}$ is also given as input, where $N$ denotes the number of total AE classes.
For example, to extract the AE sounds of the $n$-th AE class, $\mathbf{o}$ is a one-hot vector, where the element corresponding to the target AE class is set to $o_n=1$ and others are set to $o_j=0 \hspace{2.0mm} \forall j \neq n$.
Given observed mixture $\mathbf{y}$ and target-class vector $\mathbf{o}$ as inputs, the time-domain waveform of target AE class $\hat{\mathbf{x}}$ is directly estimated by the sound extraction network, as follows:
\vspace{-2mm}
\begin{align}
\hat{\mathbf{x}} = \text{DNN}(\mathbf{y}, \mathbf{o}),
\end{align}
where $\text{DNN}(\cdot)$ is the non-linear transformation of a deep neural network (DNN).
Motivated by the success of time-domain speech separation/extraction frameworks~\cite{luo2019conv,delcroix2020improving}, we adopt Conv-TasNet architecture, which demonstrated remarkable speech separation performance.
Thus, $\text{DNN}(\cdot)$ mainly consists of stacked dilated convolution blocks.

Figure~\ref{fig:network} shows a schematic diagram of the network architecture of our proposed AE sound selector network, i.e., a Sound Selector, which consists of an AE-class embedding layer and a sound extraction network.
Given $\mathbf{y}$ as input, the bottom block of the sound extraction network generates an intermediate representation of observed mixture $\mathbf{H} = \{ \mathbf{h}_{1}, \cdots, \mathbf{h}_{F} \}$, where $\mathbf{h}_{f} \in \mathbb{R}^{D \times 1}$ denotes the feature at $f$-th frame, $F$ is the total number of frames and $D$ is the dimension of the feature space.
In parallel, given $\mathbf{o} \in \mathbb{R}^{N \times 1}$ as input, an AE-class embedding layer generates target-class embedding $\mathbf{c} \in \mathbb{R}^{D \times 1}$, which provides an encoded representation of the target AE class.
These intermediate representations $\mathbf{H}$ and $\mathbf{c}$ are combined into an integrated representation $\mathbf{Z} = \{ \mathbf{z}_{1}, \cdots, \mathbf{z}_{F} \}$, which is passed to the upper blocks of the sound extraction network to output only the sounds from the target AE class. 
By integrating the target-class embeddings $\mathbf{c}$ into the sound extraction network, the network behavior can be adapted to extract the target AE sounds.

There are several ways that could be used to perform the above integration procedure.
In this paper, motivated by the success in speaker adaptation~\cite{samarakoon2016subspace} and target speech extraction~\cite{zmolikova2019speakerbeam}, we adopt an elementwise product-based integration, such as $\mathbf{z}_{f} = \mathbf{h}_{f} \odot \mathbf{c} \hspace{2.0mm} \forall f$.

\subsection{Extension to multi-class simultaneous extraction}
\label{sec:simultaneous}

We above described AE sound selection for a single AE class.
However, the Sound Selector is required to extract arbitrary sounds desired by a user depending on the situation.
This can include sounds from multiple AE classes.

To extract the sounds of multiple AE classes, we could iteratively apply the extraction module for all the target classes and then sum up the extracted signals to reconstruct the desired output.
However, such an iterative approach naturally increases the computational cost as the number of target class increases.

To alleviate the above issue, we propose a multi-class simultaneous extraction scheme that extracts multi-class AE sounds in a one-pass manner and makes the computational cost constant as the number of target classes increases.
In a multi-class selection setting, the network output remains a single signal, which includes all the sounds belonging to the target classes.

In this work, we hypothesize that the network behavior can be adapted to extract not only a single target AE class but also an arbitrary combination of multiple target AE classes.
To inform the sound extraction network about the multiple AE classes, we compute target-class embedding $\mathbf{c}$, as follows:
\vspace{-1.0mm}
\begin{align}
\mathbf{c} = \mathbf{W} \mathbf{o} = \sum_{n=1}^{N} o_{n} \mathbf{e}_{n},
\end{align}
where $\mathbf{W} = [\mathbf{e}_{1}, \cdots, \mathbf{e}_{n}, \cdots, \mathbf{e}_{N}]$ is a set of trainable weight parameters and $\mathbf{e}_{n}$ is an AE-class embedding of the $n$-th AE class.
This formalization corresponds to having a target-class vector $\mathbf{o}$ set to a $n$-hot vector, where the $n$ elements that correspond to the target AE classes are $1$ and the others are $0$.

\subsection{Training procedure}

We assume that a set of input and target features $\{ \mathbf{y}, \mathbf{o},$ $\{ \mathbf{x}_{n} \}_{n=1}^{N} \}$ is available for training the model, where $\mathbf{x}_{n} \in \mathbb{R}^{T}$ is the target sound signal of the n-th AE class. In general, the number of AE sounds in the mixture is lower than the number of total AE classes $N$.
We thus set the AE sounds of classes $\mathbf{x}_{n}$ that do not exist in the mixture at zero signals.

To realize the proposed multi-class simultaneous extraction, we dynamically generated target-class vector $\mathbf{o}$, where one or more elements are $1$ and the others are $0$, and also generated the corresponding reference signal, as follows:
\begin{align}
\vspace{-1.5mm}
\mathbf{x} = \sum_{n=1}^{N} o_{n} \mathbf{x}_{n}.
\label{eq:ref}
\end{align}

To retain the scale information of the AE sounds, we adopted the scale-dependent signal-to-noise ratio (SNR)~\cite{roux2019sdr} as the training objective.
The SNR loss $\mathcal{L}$ is expressed as follows:
\begin{align}
\mathcal{L} = 10 \log_{10} \biggl( \frac{\| \mathbf{x} \|^{2}}{\| \mathbf{x} - \hat{\mathbf{x}} \|^{2}} \biggr)
\Leftrightarrow -10 \log_{10} \bigl( \| \mathbf{x} - \hat{\mathbf{x}} \|^{2} \bigr),
\end{align}
where $\hat{\mathbf{x}}$ denotes the estimate of the target signal computed from $\mathbf{y}$ and $\mathbf{o}$.
For the baseline PIT-based separation, we use the equivalent log mean squared error (MSE) criterion to enable training when the targets are zero signals.

\subsection{Sound removal problem}

The sound removal problem is the opposite of the sound selection problem.
Sound selection aims to extract the sounds of the selected AE classes, while sound removal aims to suppress the sounds of the selected AE classes from the observed mixture.

We consider two variants to realize a sound removal mechanism: 1) direct estimation and 2) indirect estimation.
In the former scheme, we build a sound removal network by changing the reference signal in Eq~\eqref{eq:ref} to removal target $\mathbf{x} = \mathbf{y} - \sum_{n=1}^{N} o_{n} \mathbf{x}_{n}$.
In the latter scheme, we use the Sound Selector to extract the sounds to suppress and generate the removal output as $\hat{\mathbf{x}} = \mathbf{y} - \hat{\mathbf{x}}^{\text{Sel.}}$, where $\hat{\mathbf{x}}^{\text{Sel.}}$ denotes the estimate by the Sound Selector.

\section{Related work}

We base our proposed AE Sound Selector on target speech extraction framework~\cite{zmolikova2019speakerbeam,zmolikova2017speaker}, which outputs only a target speech given the target speaker information.
The advantage of this framework is that it can directly optimize the source extraction objective without requiring knowledge of the maximum number of sources in the mixture.
In this work, we introduce learned class embeddings instead of source embeddings that are derived from sound examples~\cite{zmolikova2019speakerbeam,zmolikova2017speaker}, which allow the extraction of AEs, depending on user-specified target classes.
We also extend the approach so that it can simultaneously extract AEs from multiple classes, unlike \cite{zmolikova2019speakerbeam,zmolikova2017speaker} that handle only a single target.


Combination of AES and AED has been investigated for improving AED performance for polyphonic sounds~\cite{heittola2011sound,gemmeke2013exemplar}.
For example, in recent DCASE challenges, non-negative matrix factorization (NMF)-based separation~\cite{lee1999learning,smaragdis2007supervised} was often used as a front-end for AED, e.g., \cite{komatsu2016acoustic,jeon2017nonnegative}.
Besides, \cite{tzinis2020improving} investigated an AES and AED combination to improve separation performance but not to realize AE sound selection.
It derived an iterative PIT-based separation framework, that uses the AED class posteriors of all the separated signals as auxiliary inputs to a second separation pass.
This approach inherits the limitations of the PIT-based separation frameworks, i.e., where the maximum number of sources in the mixture must be fixed.


Quite recently, \cite{kong2020source} investigated using dataset with weak-labels for AES.
In concurrent and independent work, they proposed a similar framework for extracting sounds in a mixture based on condition vector derived from an AED and evaluated their system on mixtures of 2 classes.
Compared to \cite{kong2020source}, we focus on the AE sound selection/removal and investigate explicit extraction of multiple AE classes simultaneously for both training and inference.
Besides, we evaluate our proposed scheme for extraction of up to four AE classes simultaneously on mixtures of up to seven AE classes.
In future work, we will include an investigation of using weak-labels for training~\cite{kong2020source,pishdadian2020learning}.






\section{Experiments}

\subsection{Dataset}

To evaluate the effectiveness of our proposed method, we created  datasets of simulated sound event mixtures based on the Freesound Dataset Kaggle 2018 corpus (FSD)~\cite{fonseca2018general}, which contains audio clips from 41 diverse AE classes, such as human sounds, object sounds, musical instruments, etc~\cite{fonseca2018general}.
We included stationary background noise to the mixtures using noise samples from the  REVERB challenge corpus (REVERB)~\cite{kinoshita2016summary}.
We created two datasets: 1) three-class mixtures (Mix 3) and 2) three to five-class mixtures (Mix 3-5).
The mixtures in Mix 3 contain the AEs of three classes within an utterance, and the mixtures in Mix 3-5 contain the AEs of three, four, or five classes.

We generated six-second mixtures by randomly extracting six audio clips of 1.5 to 3 seconds from the FSD corpus and pasting (adding) them to random time-positions on top of the six-second background noise.
With this way of constructing the data, different sounds from the same AE class can occur up to twice per mixture.
We created sound event mixtures by utilizing Scaper's functionality~\cite{salamon2017scaper}, ``generate\_from\_jams($\cdot$)'', which is widely used to create dataset in the SED community, where {\it ref$\_$db} was set at -50 and {\it snr} was set randomly between 15 and 25 dB for each foreground event.
In this experiment, we downsampled the sounds to 8 kHz to reduce the computational and memory costs.


The training and development sets consist of 50,000 and 10,000 mixtures, respectively.
The audio clips for these sets were randomly selected from the {\it training} sets in the FSD and REVERB corpora.
The test set consists of 10,000 mixtures based on the {\it test} set in the FSD and REVERB corpora.
For each utterance, we also randomly generated target-class vector $\mathbf{o}$ that represents the desired AE classes.

\begin{table}[t]
  \caption{Baseline SDR [dB] for mixture signals.}
  \vspace{-2mm}
  \label{tab:mixture}
  \centering
  \scalebox{0.9}{
  \begin{tabular}{ c c c c c c c }
    \toprule
     & \textbf{\# class for} & \multicolumn{3}{c}{\textbf{\# class in Mix.}} & \\
    \textbf{Dataset} & \textbf{Sel.} & \textbf{3} & \textbf{4} & \textbf{5} & \textbf{mean} \\
    \midrule
    Mix 3 & 1 & -3.6 & - & - & -3.6 \\
    \midrule
    Mix 3-5 & 1 & -3.3 & -5.9 & -7.2 & -6.0 \\
     & 2 & 3.5 & 0.0 & -2.1 & -0.2 \\
     & 3 & 21.8 & 5.9 & 2.0 & 6.3 \\
    \bottomrule
  \end{tabular}
  }
  \vspace{-3mm}
\end{table}


\subsection{Configurations}

For all the experiments, we adopted the Conv-TasNet-based network architecture, which consists of stacked dilated convolution blocks.
By following the notations of \cite{luo2019conv}, we set the hyperparameters as follows: N = 256, L = 20, B = 256, H = 512, P = 3, X = 8, and R = 4.
We also set the dimension of embedding layer $D$ at 256.
For the integration layer, we adopted element-wise product-based integration and inserted it after the first stacked convolution block (Figure~\ref{fig:network}).

We adopted the Adam algorithm~\cite{kingma2015adam} for optimization with an initial learning rate of 0.0005 and used gradient clipping~\cite{pascanu2013difficulty}.
We stopped the training procedure after 200 epochs.

As the evaluation metrics, we used the scale-invariant signal-to-distortion ratio (SDR) of BSSEval~\cite{vincent2006performance}.
In the experiment, we conducted three types of evaluations for 1) single-class selection, and multi-class selection with 2) two classes, and 3) three classes. 
For each mixture, three AE classes $\{ n_{1},$ $n_{2}, n_{3} \}$ were pre-defined.
For the multi-class ($I$-class) selection task, the reference signal for the SDR calculation is given by $\mathbf{x} = \sum_{i=1}^{I}  \mathbf{x}_{n_{i}}$, where $I$ is the number of target AE classes, i.e., $I \in \{ 1, 2, 3 \}$ in this experiment.

Table~\ref{tab:mixture} shows the average SDR scores of the mixture signals, which are used as the baseline to compute the SDR improvement scores.
Here, ``\# class for Sel.'' denotes the number of AE classes, i.e., $I$, defined as the selection target, and ``\# class for in Mix.'' denotes the subsets of the evaluation dataset with three, four, or five AE classes in the mixture.


\subsection{Results}

\subsubsection{Evaluation: single-class selection}

In the first experiment, we evaluated the single-class extraction task on Mix 3 and Mix 3-5 tasks.
Table~\ref{tab:single} shows the SDR improvement of the PIT-based system and our proposed Sound Selector.
For a fair comparison, we employed a similar network architecture for both the PIT-based and our proposed systems.
With the PIT-based system, we assumed oracle target class selection (OS), which selects one of the output sources that shows the highest SDR score as the target AE sound.
The PIT-based networks have three and five output channels for the Mix 3 and Mix 3-5 tasks, respectively.
They are trained to output zero signals when the number of sources in the mixture is fewer than the output channels~\cite{yoshioka2018multi}.

\begin{table}[t]
  \caption{SDR improvement [dB] for single-class selection.}
  \vspace{-2mm}
  \label{tab:single}
  \centering
   \scalebox{0.9}{
  \begin{tabular}{ c c c c c c }
    \toprule
    & & \multicolumn{3}{c}{\textbf{\# class in Mix.}} & \\
    \textbf{Method} & \textbf{Dataset} & \textbf{3} & \textbf{4} & \textbf{5} & \textbf{mean} \\
    \midrule
    PIT + OS & \text{Mix 3} & 12.0 & - & - & 12.0 \\
    Sound Selector & \text{Mix 3} & 11.2 & - & - & 11.2 \\
    \midrule
    PIT + OS\footnotemark & \text{Mix 3-5} & 9.6 & 9.1 & 8.5 & 9.0 \\
    Sound Selector & \text{Mix 3-5} & 8.8 & 10.5 & 11.1 & 10.5 \\
    \bottomrule
  \end{tabular}
  }
  \vspace{-0mm}
\end{table}
\footnotetext{We observed that PIT-based training with SNR loss failed to output the correct number of sources in the mixture when trained on the Mix 3-5 set.
Additional investigations with alternative training objectives or curriculum learning may be required to prevent this issue.
}

Table~\ref{tab:single} shows that the SDR score of the Sound Selector is slightly lower than the PIT+OS for Mix 3 task, but the proposed Sound Selector stably worked better than the PIT+OS for a more realistic Mix 3-5 task, i.e., where the number of classes in the mixture is variable.
These results demonstrate that the proposed architecture can jointly solve the AES and AED tasks.

Note that the scores of the PIT+OS would correspond to the upper bound performance with the oracle AE classifier.
In practice, performing AED on top of the separation would be difficult due to potential distortion or remaining undesired sounds in the signals\footnote{Even for clean signals, the MAP@3 score is about 95 \% on the same FSD corpus~\cite{fonseca2018general} and the top-1 accuracy would be even lower.}.
Therefore, if we adopt the actual AE classifier, the scores of the PIT+OS will certainly be degraded.
Evaluation with such a real classifier would be important future work.

\begin{table}[t]
  \caption{SDR improvement [dB] for multi-class selection.}
  \vspace{-2mm}
  \label{tab:multi}
  \centering
  \scalebox{0.9}{
  \begin{tabular}{ c c c c c c c }
    \toprule
     & \textbf{\# class for} & \multicolumn{3}{c}{\textbf{\# class in Mix.}} & \\
    \textbf{Method} & \textbf{Sel.} & \textbf{3} & \textbf{4} & \textbf{5} & \textbf{mean} \\
    \midrule
    Iterative & 2 & 4.4 & 6.7 & 8.0 & 6.9 \\
    Simultaneous & 2 & 5.1 & 6.8 & 7.7 & 6.9 \\
    \midrule
    Iterative & 3 & -11.0 & 2.7 & 5.2 & 2.0 \\
    Simultaneous & 3 & -4.9 & 3.9 & 5.0 & 3.3 \\
    \bottomrule
  \end{tabular}
  }
  \vspace{-3mm}
\end{table}

\subsubsection{Evaluation: multi-class selection}

In the second experiment, we evaluated the multi-class extraction task on the Mix 3-5 task.
Table~\ref{tab:multi} shows the SDR improvement of the iterative and simultaneous extraction schemes.
Here, ``Simultaneous'' denotes the proposed multi-class simultaneous extraction scheme, as described in Section~\ref{sec:simultaneous}.
The ``Iterative'' scheme iteratively applies single-class extraction for each of the target AE classes and sums up all of the extracted sounds.

Table~\ref{tab:multi} shows that the ``Simultaneous'' extraction scheme works comparably or better than the ``Iterative'' scheme, regardless of its lower computational cost.
These results suggest that the proposed AE selection concept could be achievable with a constant computational cost.

We observe that the SDR improvement is negative when both ``\# class for Sel.'' and ``\# class for Mix.'' are three.
This is due to the high baseline SDR (21.8 dB in Table~\ref{tab:mixture}), because the reference signals closely resemble the mixtures in this setting.
Note that for the ``Simultaneous'' scheme, although SDR is degraded by 4.9 dB, the absolute SDR remains high (16.9 dB), showing that the output signal is not significantly distorted.

Due to space limitations, we are unable to report the detailed results on the sound removal experiments, but we confirmed that both direct and indirect estimation schemes performed equivalently with SDR improvement of about 6 dBs.

\subsubsection{Analysis of generalization capability}

\begin{figure}[t]
  \centering
  \includegraphics[width=\linewidth]{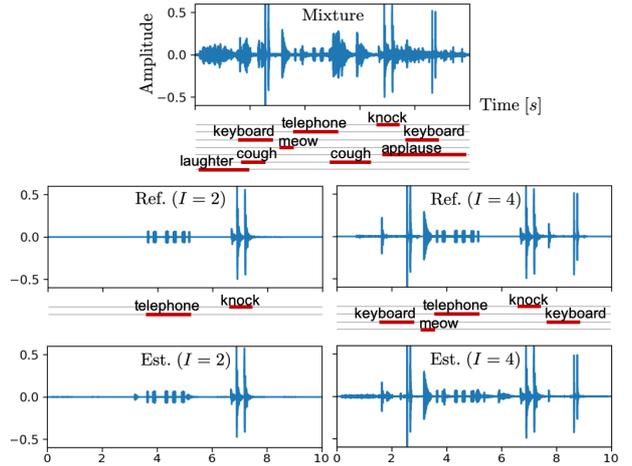}
  \vspace{-5mm}
  \caption{Mixture with AEs from 7 AE classes, the reference signals (Ref.), and the signals estimated with the proposed Sound Selector (Est.) for 2  (left) and 4 (right) target AE classes.}
  \label{fig:example}
  \vspace{-5mm}
\end{figure}

Finally, we explored the generalization capability of our approach to conditions unseen during training, i.e., longer mixtures with more AE classes and more target classes for selection.
We created an additional test set consisting of 200 home office-like mixtures of ten seconds containing AEs from seven classes.
The target AE classes are set for all mixtures to ``knock, telephone'' for two-class ($I=2$) and ``knock, telephone, keyboard, meow'' for four-class ($I=4$) cases.

Figure~\ref{fig:example} shows an example of sound selection for a AE mixture\footnote{Audio examples of the proposed sound selection/removal system will be available online~\cite{example}}, where ''Ref`` and ``Est'' denote reference and estimated signals, respectively.
The results were obtained with the multi-class simultaneous extraction scheme.
From the figure, we confirmed that our proposed system successfully extracts the sounds of the multiple target  AE classes, even if seven-class mixtures and four-class simultaneous extraction were not included in the training stage.

Figure \ref{fig:example} shows only one example, but we observed that the average SDR improvements on this set are 8.5 dB for two-class and 5.3 dB for four-class cases.
This result suggests that the proposed framework has the potential to generalize to unseen conditions. 

\section{Conclusions}

In this paper, we introduced two novel concepts, i.e., AE sound selection and removal problems. Solving these problems would open a path toward hearable devices that allows the selection of AEs to which we want to listen.
We proposed a sound selection neural network, i.e., Sound Selector, which exploits a $n$-hot representation of the user-specified target AE classes to simultaneously output sounds from these classes.
Experimental results showed that the proposed Sound Selector successfully extracts the multiple AE sounds simultaneously and it has the potential to generalize to an unseen number of classes in the mixture and to the extraction of an arbitrary number of classes.

Future work will include investigations using a larger dataset with more AE classes, which would be essential for further evaluating the generalization capability of the proposed method, as well as extension to online processing and training with weak-labels.

\clearpage

\bibliographystyle{IEEEtran}

\bibliography{mybib}

\end{document}